\documentclass[aps, twocolumn,
superscriptaddress,showpacs,showkeys,amsmath,amssymb,floatfix]{revtex4}
\usepackage{color}
\usepackage{graphicx}
\usepackage{epsfig}
\usepackage{dcolumn}
\usepackage{bm}
\usepackage{mathtools}
\usepackage{amssymb}
\usepackage{bmpsize}
\usepackage{calrsfs}
\DeclareMathAlphabet{\pazocal}{OMS}{zplm}{m}{n} 
\usepackage{amsmath}
\usepackage{subfigure}
\newcommand{\be}{\begin{equation}}
\newcommand{\ee}{\end{equation}}
\newcommand{\bea}{\begin{eqnarray}}
\newcommand{\eea}{\end{eqnarray}}

\newcommand{\hi}{{\hat i}}
\newcommand{\hu}{{\hat u}}

\newcommand{\tK}{{\tilde K}}

\newcommand{\tF}{{\tilde F}}

\newcommand{\tg}{{\tilde g}}

\newcommand{\tM}{{\tilde M}}

\newcommand{\tk}{{\tilde k}}
\newcommand{\tw}{{\tilde w}}

\newcommand{\pro}{\partial}

\newcommand{\bfA}{{\mathbf A}}

\newcommand{\bfr}{{\mathbf r}}

\newcommand{\bfw}{{\mathbf w}}

\newcommand{\ba}{\begin{array}}
\newcommand{\ea}{\end{array}}

\newcommand{\nn}{\nonumber}

\newcommand{\Ab}{\pazocal A}

\begin{document} 
\title{Inherent color symmetry in quantum Yang-Mills theory}
\author{Dmitriy G. Pak}
\affiliation{Institute of Theoretical Physics, Chinese Academy of Sciences, Beijing 100190, China}
\affiliation{School of Education, Bukkyo University, Kyoto 603-8301, Japan}
\author{Rong-Gen Cai}
\affiliation{Institute of Theoretical Physics, Chinese Academy of Sciences, Beijing 100190, China}
\author{Takuya Tsukioka}
\affiliation{School of Education, Bukkyo University, Kyoto 603-8301, Japan}
\author{Pengming Zhang}   
\affiliation{School of Physics and Astronomy,
 Sun Yat-sen University, Zhuhai, 519000, China}
\author{Yu-Feng Zhou}   
\affiliation{Institute of Theoretical Physics, Chinese Academy of Sciences, Beijing 100190, China}
\begin{abstract}
We  present the basic non-perturbative structure of the space of classical dynamical solutions and 
corresponding one particle quantum states in $SU(3)$ Yang-Mills theory. It has been demonstrated that
the Weyl group of $su(3)$ algebra plays an important role in constructing non-perturbative solutions 
and leads to profound changes in the structure of the classical and quantum Yang-Mills theory. 
We show that the Weyl group as a non-trivial color subgroup of $SU(3)$ admits singlet irreducible representations
on a space of classical dynamical solutions which lead to strict concepts of one particle quantum states for gluons and quarks.
The Yang-Mills theory is a non-linear theory and, in general, it is not possible to construct a Hilbert space of classical solutions
and quantum states as a linear vector space, so, usually, a perturbative approach is applied.
We propose a non-perturbative approach based on
Weyl symmetric solutions to full non-linear equations of motion and construct a full space of dynamical solutions representing
an infinite but countable solution space classified by a finite set of integer numbers. It has been proved that
the Weyl singlet structure of classical solutions provides the existence of a stable non-degenerate vacuum 
 which serves as a main precondition of the color confinement phenomenon. 
 Some physical implications in quantum chromodynamics are considered.
\end{abstract}
\maketitle
  
     \section{Introduction}
     
     Construction of a strict non-perturbative quantum theory of strong interaction on a basis of  $SU(3)$
quantum Yang-Mills theory represents one of the most fundamental problems in theoretical physics \cite{Mill2000}. 
A primary task in constructing a quantum field theory is to determine 
a space of classical dynamical solutions defining the Hilbert space of one particle quantum states. In quantum 
chromodynamics (QCD) due to the presence of the confinement phenomenon the single color gluons and quarks are 
not observed, so such states should be excluded from the spectrum of physical quantum states.
It is observed \cite{wilson1974, kogut-susskind1975} that origin of color confinement is conditioned by 
existence of a non-degenerate vacuum, which must be color invariant \cite{polyakov77} and
described by Abelian colorless gluons in the confinement phase \cite{thooft81}.
With this one encounters a problem with construction of singlet classical solutions which could provide
a color invariant and non-degenerate vacuum, and consistent definition of one particle quantum states.  
The problem is related with a simple mathematical fact, that color group $SU(3)$ and its continious subgroups do not admit non-trivial singlet
irreducible representations. Selection of a vacuum belonging to the reducible representation leads to degenerate vacuum and spontaneous
color symmetry breaking which is incompatible with the color confinement. On the other hand,
after fixing the gauge color symmetry one needs some kind of residual color symmetry to find solutions
with inherent symmetry which would provide non-degenerate color invariant vacuum and
color attributes of fundamental particles, gluons and quarks.
It is surprising, that the problem of searching solutions with inherent color symmetries in non-Abelian $SU(3)$ Yang-Mills
theory can be resolved due to 
the presence of the Weyl symmetry group of $SU(3)$ which is the 
only color symmetry which survives after gauge fixing. 
The most important feature of the Weyl group is that it has singlet irreducible representations which allow to construct
a deepest stable non-degenerate vacuum of QCD and describe a full space of singlet dynamical solutions
leading to one particle quantum states.

In this Letter we present the basic construction of a space of Weyl symmetric classical stationary solutions 
realizing singlet irreducible representations of the Weyl group. Due to non-linear structure of Yang-Mills
equations the space of solutions can not be supplied with a linear vector space structure as one has in Abelian
theories like the electrodynamics. This leads to the absence of linear superposition rule in the theory, as a consequence,
the space of quantum states does not represents a Hilbert vector space and it is not possible to define a complete set of basis solutions.
Nevertheless, we demonstrate that all configuration space of regular finite energy 
stationary solutions is infinite and countable. This allows to construct a full space of discrete one particle quantum states
classified by a finite set of integer quantum numbers. 
Note that direct canonical quantization of stationary solutions leads to quantum states
which do not represent directly physical observable quantities. Due to generation of a non-trivial vacuum and 
non-zero vacuum gluon and quark condensates one has to take into account the vacuum polarization effect which
leads to effective interaction of the singlet quantum states with vacuum condensates. 
We show that such interaction implies formation 
of localized bound states representing physical observables in QCD, glueballs and mesons.

\section{Ansatz for singlet Weyl symmetric solutions}

Recently an ansatz for classical stationary solutions symmetric under Weyl group transformations has been proposed.
It has been proved that Weyl symmetric solutions are stable against quantum gluon fluctuations, and the 
Abelian type solution provides a stable non-degenerate QCD vacuum \cite{plb2018} resolving a long-standing problem of 
vacuum instability in QCD \cite{savv, N-O}.
In this section we prove that solutions defined by the Weyl symmetric ansatz possess a remarkable property:
they describe only singlet irreducible representations of the Weyl group (classification of Weyl group representations is briefly described
in the supplemental material \cite{SM2}).

   We start  with a standard Lagrangian for $SU(3)$ Yang-Mills theory
    $(\mu,\nu=0,1,2,3; a=1,2,...,8)$
   \bea
{\cal L}_{YM} = -\dfrac{1}{4} F_{\mu\nu}^a F^{a\mu\nu}. \label{Lagr0}
\eea
    Our primary goal is to construct a space of regular stationary singlet Weyl symmetric solutions leading to one 
    particle quantum states after standard quantization.
         After minor changing notations in the ansatz proposed in \cite{plb2018} we define first an extended Weyl symmetric 
ansatz by setting non-vanishing components of the gauge potential $A_\mu^a$ corresponding to $I,U,V$ type subgroups 
$SU(2)$ \cite{neeman99} as follows
\begin{align} 
I:~~A_t^2&=K_0, &A_r^2&=K_1, & A_\theta^2&=K_2, & A_\varphi^1&=K_4,\nn \\
U:~A_t^5&=-Q_0, & A_r^5&=-Q_1, & A_\theta^5&=-Q_2, & A_\varphi^4&=Q_4, \nn \\
V:~A_t^7&=S_0, & A_r^7&=S_1, & A_\theta^7&=S_2, & A_\varphi^6&=S_4,\nn \\
{\Ab}_\varphi^p&=A_\varphi^\alpha r_\alpha^{\,p}, & 
A_\varphi^3&=K_3, & A_\varphi^8&= K_8, \label{SU3DHN} 
\end{align}
where $r_\alpha^{\,p}$ $(\alpha=3,8)$ are root vectors
$\bfr^{1}=(1,0)$, $\bfr^{2}=(-1/2,\sqrt 3/2)$, $\bfr^{3}=(-1/2,-\sqrt 3/2)$,
 (index $p=1,2,3$ denotes $I,U,V$ sectors).
To fix residual local color symmetry $U(1)$ in $I,U,V$ sectors of the Yang-Mills Lagrangian, we 
add Lorenz type gauge fixing terms ${\cal L}_{\rm gf}$ to the original Yang-Mills Lagrangian ${\cal L}_{YM}$
 \be
{\cal L}_{\rm gf}=-\dfrac{1}{2}\sum_{a=2,5,7}(\pro_t A_t^a-\pro_r A_r^a-\dfrac{1}{r^2} \pro_\theta A_\theta^a)^2. 
\label{gfterm}
\ee
The gauge fixing procedure is a necessary step during quantization which removes all unphysical pure gauge field degrees of freedom.
The gauge fixing Lagrangian (\ref{gfterm}) fixes not only the local gauge symmetry, but also the global color symmetry 
$SU(3)$. So the ansatz
is symmetric only under the Weyl group transformations, providing only Weyl representations for solutions.

It is suitable to choose representation of the Weyl group as a symmetric group $S_3$ 
acting on fields $A_\mu^a$ in $I,U,V$ sectors by permutations. So, the original octet $A_\mu^a$ realizes 
the eight dimensional vector reducible representation $\Gamma_8$ of the Weyl group. 
One can define an ansatz describing only singlet Weyl representations by imposing additional constraints providing consistency
of the full ansatz with all Yang-Mills equations of motion ($\hi=0,1,2$)
\bea
Q_\hi&=&S_\hi=K_\hi,~~~~~~K_4+Q_4+S_4=0, \nn \\
Q_4\!\!&=&\!\!\Big (-\frac{1}{2}+\frac{\sqrt 3}{2}\Big ) K_4, \quad
S_4= \Big (-\frac{1}{2}-\frac{\sqrt 3}{2}\Big ) K_4, \quad \nn \\
K_3&=& -\dfrac{\sqrt 3}{2}K_4,~~~~K_3=K_8, \label{reduction1}
 \eea
The ansatz (\ref{SU3DHN}, \ref{reduction1}) extracts three singlet representations $\{\Gamma_1\}_\hi$ of $S_3$
 for fields $(K_\hi,Q_\hi,S_\hi)$ in color subspace spanned by generators $\{T^{2,5,7}\}$. 
 Constraints (\ref{reduction1}) imply
that representation acting on $(K_4, Q_4,S_4)$ is isomorphic to $S_3$ representation defined on $I,U,V$ fields
 ${\Ab}_\varphi^p$ (\ref{SU3DHN}). 
 Let us consider eigenvalues of a Lie algebra valued Abelian field,
 $\bfA_\varphi = (A_\varphi^3 T^3,~A_\varphi^8 T^8)$, acting in adjoint representation 
 in the Cartan basis. Due to the constraint 
$K_3=K_8$, (\ref{reduction1}), one can find
 \bea 
&&[\bfA_\varphi^p, T_+^p]=K_3 \bfr^p T_+^p, \label{K3root}
\eea
where the eigenvalues $K_3 \bfr^p$ by module $K_3$ define color charges of $I,U,V$-components
of the Abelian field. A total color charge of the Abelian field ${\Ab}_\varphi^p$ is zero, and eigenvalues $K_3 \bfr^p$ 
match the symmetric system of three root vectors $\bfr^p$ with a common field factor $K_3$.
The Weyl group is defined as a symmetry group of root system, so that
 $K_3 \bfr^p$ realize a singlet symmetry representation $\Gamma_1^s$ \cite{SM2},
 and $K_3$ (or $K_4$ equivalently), representing a fixed point in the configuration space of fields under Weyl transformations,
is a Weyl invariant Abelian field. 

Applying ansatz  (\ref{SU3DHN}, \ref{reduction1}) to $SU(3)$ Yang-Mills Lagrangian with gauge fixing terms, (\ref{gfterm}),
one can write the total Lagrangian in an explicit Weyl symmetric form
 \bea
&&{\cal L}_{tot}^{Weyl}={\cal L}_{YM}+\sum_{\text{\tiny {I,U,V}}}{\cal L}_{g.f.}^{I,U,V}\nn \\
&&=\sum_{p}\Big \{  -\dfrac{1}{3} (\pro_\mu \Ab_\nu^p)^2-|D_\mu^p W_\nu^p|^2\nn \\ 
&&- \dfrac{9}{4} \Big (
 (W^{*p\mu} W^p_\mu)^2-(W^{*p\mu} W^{*p}_\mu) (W^{p\nu} W^p_\nu) \Big )\Big \},  \label{Lweyl}
 \eea
 with
 \bea
&&W^{I}_{\mu}= \dfrac{1}{\sqrt 2} (A_\mu^1 + i A_\mu^2), ~~~~W^{U}_{\mu}=
                                                 \dfrac{1}{\sqrt 2} (A_\mu^4- i A_\mu^5),\nn \\
&&W^{V}_{\mu}=\dfrac{1}{\sqrt 2}(A_\mu^6+i A_\mu^7),~~D_\mu^p=\pro_\mu+i A_\varphi^\alpha r_\alpha^{\,p}.\nn
\eea
 Note, that in general, a Weyl symmetric Lagrangian does not guarantee existence of singlet Weyl symmetric solutions.
For instance, a one-loop effective potential with two Abelian fields $A_\mu^{3,8} \in \Gamma_2$ can be written
in a manifest Weyl symmetric form, however, it has two degenerate vacuums.
Contrary to this, an effective potential with one Weyl symmetric Abelian field under condition $A_\varphi^3=A_\varphi^8$
has one deepest non-degenerate vacuum \cite{flyvb,mpla2006}. The Lagrangian ${\cal L}_{tot}^{Weyl}$ can be rewritten
 in terms of four independent fields $K_\mu$ (world index $\mu=0,1,2,3$ denotes space-time 
 coordinates $(t,r,\theta,\varphi)$
\bea
&&{\cal L}_{\rm red}(K)= \nn\\
&&\dfrac{3}{2r^2}\Big [r^2(\pro_t K_1 -\pro_r K_0)^2-(\pro_\theta K_1)^2+(\pro_\theta K_0)^2\Big ]\nn\\
&&+\dfrac{3}{2 r^2} \Big [ \pro_t K_2 (\pro_t K_2- \pro_\theta K_0) - \pro_r K_2(\pro_r K_2-\pro_\theta K_1) \Big ]\nn \\
&&+\dfrac{27}{16r^4 \sin^2 \theta}\Big [ r^2 ((\pro_t K_3)^2-(\pro_r K_3)^2) -(\pro_\theta K_3)^2 \Big] \nn \\
& &     -\dfrac{27}{16 r^4\sin^2 \theta} \Big [K_3^2 (K_2^2+r^2 (K_1^2-K_0^2)) \Big ].\label{Lred}
\eea
After substitution of the ansatz  (\ref{SU3DHN}, \ref{reduction1}) into Euler equations corresponding to the total Lagrangian
${\cal L}_{tot}^{Weyl}$  
equations of motion reduce to four non-degenerated second order partial differential equations and one constraint 
for four independent fields $K_\mu$ (\cite{SM2}, (5-9)).

Let us consider the structure of the space of Weyl symmetric solutions.   
It is instructive to consider first a simple case of Abelian reduced Weyl symetric ansatz (\ref{SU3DHN})
obtained by setting to zero all off-diagonal components of the gauge potential $A_\mu^a$. Non-vanishing
Abelian potentials $A_\mu^{3,8}$ define two constant Abelian chromomagnetic fields $H_{\mu\nu}^{3,8}$.
For external constant magnetic fields an explicit analytical  
expression for one-loop effective potential is known \cite{flyvb, mpla2006}. The effective potential describes the vacuum energy
which can be considered as a function of two number parameters 
$H^{3,8}={\sqrt {(H_{\mu\nu}^{3,8})^2}}$. The potential has two degenerate vacuums located at the plane $(H^3, H^8)$:
$
H^3=H_0, H^8=0$ and at $H^3=\dfrac{1}{2} H_0, H^8=\dfrac{\sqrt 3}{2}H_0$.
Since the effective potential is invariant under reflections
$H^{3,8}\rightarrow \pm H^{3,8}$ one has six degenerate vacuums which form the Weyl sextet representation 
corresponding to the Weyl symmetry of root diagram of $SU(3)$ Lie algebra. Note that six vacuum component pairs $(H^3, H^8)$
up to the common factor $K_0$ match exactly the roots, so that the vacuum sextet belongs to two-dimensional 
reducible Weyl representation $\Gamma_2$. 
In addition, the effective potential has an absolute minimum $H^3=H^8=2^{-1/3} H_0$ which corresponds to a Weyl 
singlet representation $\Gamma_1$ due to the constraint $K^3=K^8$ in the ansatz for Weyl singlet solutions (\ref{reduction1}).
One can verify \cite{mpla2006} that local vacuums from the Weyl sextet are unstable (saddle points), and they are parameterized by an 
additional angle parameter, $\cos \theta=(H^3{\mu\nu}\cdot H^8_{\mu\nu})/H^3 H^8$. In the limit $\theta \rightarrow \pi/2$ all six local vacuums
merge into the absolute vacuum. For the absolute vacuum the Weyl invariant variables $H^p=(H^I,H^U,H^V)$ \cite{mpla2006} 
\bea
H^p=\sqrt {(H^p_{\mu\nu})^2}, ~~~~~H^p_{\mu\nu}=r^p_{\alpha} H^\alpha_{\mu\nu}, ~~~~\alpha=3,8;
\eea
take the same value,  $H_I=H_U=H_V\equiv H_0$. So that the deepest vacuum is located at the point
$(H_0,H_0,H_0)$ which represents a singlet standard representation $\Gamma_1$ of the permutation group $S_3$, i.e., the Weyl group.

Therefore, the Weyl symmetric non-singlet gluon solutions possess a higher symmetry to compare with non-Weyl 
symmetric solutions, and they form a
degenerate vacuum sextet. The Weyl symmetric singlet solution reveals a highest inherent
color symmetry which provides a deepest non-degenerate unique vacuum. This is the origin of vacuum stability 
against classical and quantum fluctuations what was proved first numerically in \cite{plb2018}.

 \section{Singlet structure of non-Abelian solutions}

Abelian solutions form a linear vector space, so the Hilbert space of the Abelian Weyl symmetric dynamical solutions 
is defined straightforward by a complete basis of transverse vector spherical harmonics of magnetic, $\vec A_{lm}^{\mathfrak{m}}$,
and electric, $\vec A_{lm}^{\mathfrak{e}}$,  type \cite{jackson}.
     Let us return back to consideration of the structure of a space of non-Abelian Weyl symmetric solutions.
Non-Abelian Weyl symmetric solutions are defined by four fields $K_\mu$ which satisfy a system of 
  non-linear partial  differential equations, and can be obtained only numerically. 
  A numeric solution of magnetic type with the lowest energy density is presented in Fig. 1
in the leading order of Fourier series decomposition 
\be
\begin{array}{rcl}
K_{1,2,4}(r,\theta,t)\!\!\!&=&\!\!\!\tilde K_{1,2,4}(M r,\theta)  \cos (M t),  \\ [2\jot]
K_0(r,\theta,t)\!\!\!&=&\!\!\!\tilde K_0(M r,\theta)  \sin(M t), 
\end{array}
\label{seriesdec}
\ee
where $M$  is a conformal mass scale parameter.

\begin{figure}[h!]
\centering
\subfigure[~]{\includegraphics[width=42mm,height=32mm]{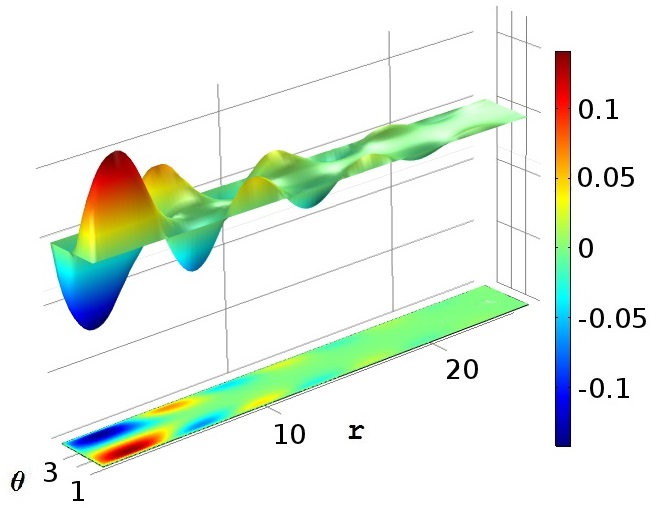}}
\hfill
\subfigure[~]{\includegraphics[width=42mm,height=32mm]{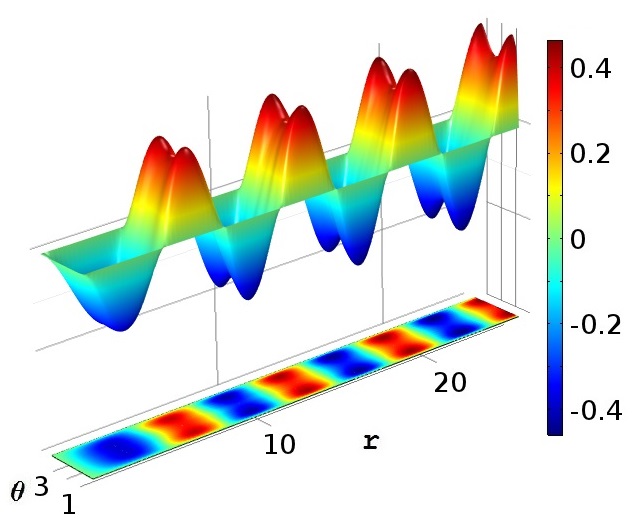}}
\hfill
\subfigure[~]{\includegraphics[width=42mm,height=32mm]{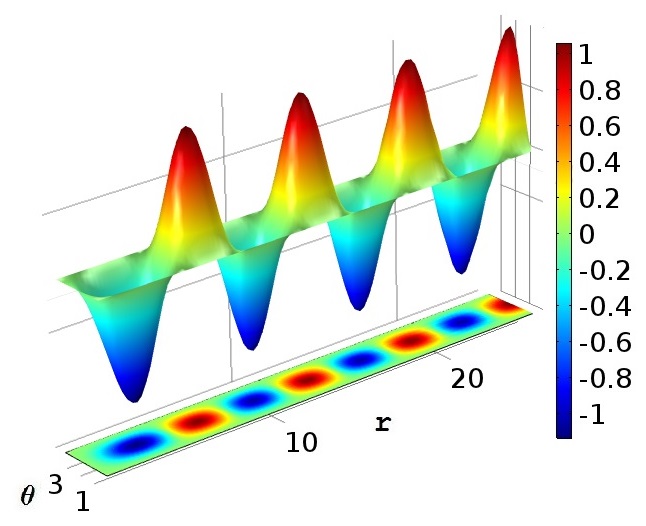}}
\hfill
\subfigure[~]{\includegraphics[width=42mm,height=32mm]{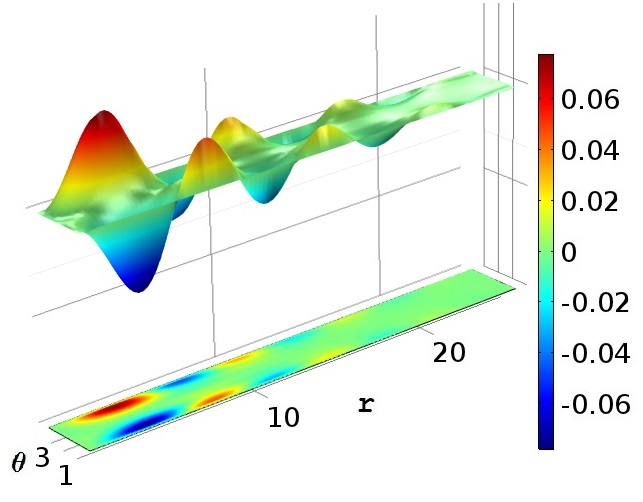}}
\hfill
\subfigure[~]{\includegraphics[width=42mm,height=22mm]{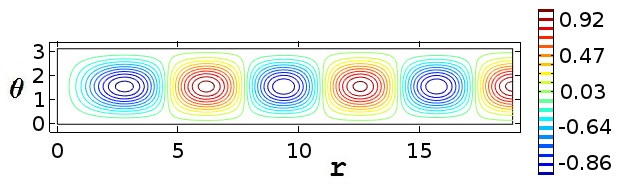}}
\hfill
\subfigure[~]{\includegraphics[width=42mm,height=24mm]{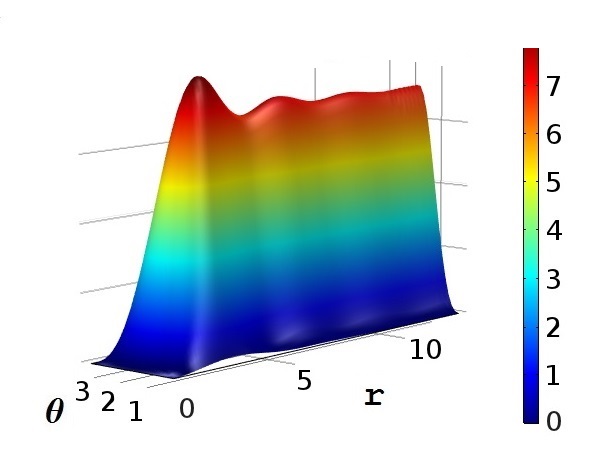}}
\caption[fig1]{Non-Abelian Weyl symmetric solution: (a) $\tK_1$;
(b) $\tK_2$; (c) $\tK_4$; (d) $\tK_0$; (e)  A contour plot of the Abelian field $\tK_4$;
(f) the time averaged energy density $r^2\sin\theta \,{\cal E}$.($g=1,M=1$).}\label{Fig1}
\end{figure}

The non-Abelian solutions reveal the Abelian dominance effect for low energy solutions.
Indeed, the Abelian numeric profile function $\tK_4(r,\theta)$, Fig. 1(c), coincides with the lowest vector harmonic 
$A_{10}^{\mathfrak{m}}$ with a high accuracy, Fig. 1(e) and TABLE I.
Moreover, one finds that a contribution of the Abelian field
to the total energy inside finite space domain is near $95\% \pm 1.5 \%$, 
which is very close to a known estimate
established in the Wilson loop functional \cite{abeldom1,abeldom3}.
\begin{table}[h]
\begin{center}
\begin{tabular}{|c|c|c||c|c|c|}
\hline
                &           &       &    &    &                    \\
~$\nu_{n1}$  & ~ $\nu_{n1}^{\rm num}$   &  ~$\nu_{n1}^{\rm exact}$ &~ $\mu_{n1} $ & ~
  $\mu_{n1}^{\rm num}$   &~  $\mu_{n1}^{\rm exact}$ \\
\hline
\hline
~$\nu_{11}$ &   2.79  &   2.74 &   $\mu_{11}$&  4.52   &  4.49  \\
\hline
~$\nu_{21}$ &   6.18  &   6.12 &   $\mu_{21}$&  7.81   & 7.73  \\
\hline
~$\nu_{31}$ &   9.37  &  9.32 &   $\mu_{31}$&  10.97   &  10.90  \\
\hline
~$\nu_{41}$ &   12.56  &   12.49 &   $\mu_{41}$&  14.14   & 14.07  \\
\hline
~$\nu_{51}$ &   15.74  &   15.64 &   $\mu_{51}$&  17.29   &  17.22  \\
\hline
\end{tabular} 
\caption{\label{tab:tc} Values of zeros and extremums of the numeric solution $\tK_4$, 
and exact values of nodes $\mu_{nl}$ and antinodes $\nu_{nl}$ of the radial part $r j_l(r)$ of the vector harmonic
$\vec A_{lm}^{\mathfrak{m}}$ ($l=1, m=0$).}
\end{center}
\end{table}
 The source of the Abelian dominance is a fact that non-Abelian solution exists only in the presence of Abelian field.
 If the Abelian field vanishes, the equations for $K_\hi$ will turn into free Maxwell equations for electric type solutions.
 The presence of Abelian dominance allows to classify all regular non-Abelian
 Weyl symmetric solutions (at least for numeric construction) using the basis of spherical harmonic 
 functions $\vec A_{lm}^{\mathfrak{m},\mathfrak{e}}$  as an initial basis for the Abelian component of 
 non-Abelian solutions.
For a given Abelian solution with selected eigenvalues $(M, J=l)$ of the energy and angular momentum operators
one defines an infinite countable set of non-Abelian solutions numerated by integer number $k=0,1,2...$ equaled to
a number of zeros of the field $\tK_2(r, \theta)$ inside interval $\theta \in (0,\pi)$ 
at a fixed point $r_0$ corresponding to any antinode $\nu_{nl}$ of the Bessel function $r j_l(r)$.
A solution with the lowest polar angle mode, $k=0$, is shown in Fig. 1. 
Solutions with higher polar modes, $k=1,2,3$, are presented in (\cite{SM2}, Figs. 1,2,3). 

The most important consequence of the ansatz (\ref{SU3DHN}, \ref{reduction1}) is that the original Yang-Mills 
Lagrangian with gauge fixing terms contains a set of four independent fields
$K_\mu$ which belong to singlet representation of the Weyl group. This implies that 
on a space of solutions one has only one Weyl  singlet four-vector field $K_\mu$ which describes only one dynamical field
degree of freedom corresponding to magnetic polarization. Note that an extended Weyl symmetric ansatz
 (\ref{SU3DHN}) describes only magnetic type solutions, electric type solutions
are defined by a different, dual ansatz  (\cite{SM2}, section III). Magnetic and electric type solutions correspond exactly 
to two possible polarizations of the original transverse gluon field $A_\mu^a$. 
Therefore, the non-Abelian Weyl symmetric singlet solution of magnetic type should contain only one
independent integration constant corresponding to the amplitude parameter
which after quantization leads to one type of creation/annihilation operator.
Indeed, a careful numeric analysis of solutions with quantum numbers $l=1, k=0,1,2,3$ 
shows that field components  $K_\mu$ for each non-Abelian solution
has only one independent normalization constant which can be assigned to the amplitude of the Abelian field $K_3$ (or $K_4$
equivalently) (\cite{SM2}, section II). Once fixed the amplitude of the Abelian field $K_4$ the amplitudes of other three
field components $K_{0,1,2}$ are uniquely determined irrespectively of their initial values 
given in initial profile functions and on boundaries in the numeric procedure \cite{SM2}. 
This implies that after quantization each regular non-Abelian Weyl symmetric solution leads to Weyl singlet one particle 
quantum state defined by given numbers $(M,J=l, k)$ in agreement with the singlet structure of the ansatz
(\ref{SU3DHN}, \ref{reduction1}).
  Note that a source of this phenomenon lies in a general restriction for solutions with a conserved energy.
Namely, the solutions must be stationary, and propagating fields $K_{2,4}$ must have the same space and time 
periodicity which provides the energy conservation in finite space domains constrained by nodes or antinodes.
{\it It is remarkable, all regular stationary Weyl symmetric solutions defined by ansatz (\ref{SU3DHN}, \ref{reduction1})
represent singlet irreducible Weyl representations classified by one conformal mass parameter $M$ and 
 integer numbers $(l,k)$.} In applications to hadron physics one has to consider solutions 
 defined in finite space domains, and the conformal parameter $M$ takes only discrete values,
 $M_{nl}=\mu_{nl} (\nu_{nl})$.
One has similar results for electric type solutions (\cite{SM2}, section III). 

\section{Weyl symmetric quark solutions}

Now we consider properties of the Weyl symmetric matter field described by the Lagrangian ${\cal L}_{\rm q}$
with one $SU(3)$ fundamental quark triplet
\bea 
{\cal L}_{\rm q}= \bar \Psi \Big[ i \gamma^\mu (\pro_\mu -\dfrac{i g}{2} A_\mu^a \lambda^a)-m\Big]\Psi.
\eea
The Euler equation for one flavor quark in the presence of gluon field $A_\mu^a$ reads
\bea
&& \Big[ i \gamma^\mu (\pro_\mu -\dfrac{i g}{2} A_\mu^a \lambda^a)-m\Big]\Psi=0, \label{eqPsi}
\eea
where the gluon field $A_\mu^a$ satisfies pure Yang-Mills equations without source
$ j^a = -\dfrac{g}{2}  \bar \Psi \gamma_\nu \lambda^a\Psi$
\bea
&&(D^\mu \vec F_{\mu\nu})^a= 0. \label{eqA}
\eea
A usual simple Abelian projection with  two independent Abelian fields $A_\mu^{3,8}$ corresponding to 
two Cartan generators leads to Dirac equations for three independent $SU(3)$ color quarks 
\bea
   \Big[ i \gamma^\mu \pro_\mu -\dfrac{i g}{2}  \sum_p \gamma^\mu A_\mu^\alpha w_\alpha^p-m\Big]\Psi_p=0,
\label{usualeq}
\eea
where $w_\alpha^p$ are the weight vectors $\bfw^p=\{(1, 1/\sqrt 3), (-1,1/\sqrt 3), (0,-2/\sqrt 3)\}$.
The equations have three independent solutions for quarks forming an irreducible color triplet in the 
fundamental representation of the color group $SU(3)$.
Note that the simple Abelian projection is not consistent with the Weyl symmetric structure defined by ansatz
(\ref{SU3DHN}, \ref{reduction1}). 
An important feature of the Weyl symmetric ansatz (\ref{SU3DHN},\ref{reduction1})
 is that it implies a non-trivial Abelian projection with one Abelian field $K_3$ located in the extended  
 color subspace spanned by generators $(T^{3,8,1,4,6})$. 
Substituting the Weyl symmetric ansatz into equation for quarks  (\ref{eqPsi}) one obtains
the following equation
\begin{align}
 &\Big[ i \gamma^\mu \pro_\mu-m+\dfrac{g}{2} \gamma^\mu  A_\mu G 
                +\dfrac{g}{2}\gamma^\mu {\hat A}_\mu Q\Big]\Psi=0 \label{GQeqns}
                \end{align}
                with color charge matrices $G$ and $Q$
                \begin{align}
 &G=\begin{pmatrix}
   \tw^1&\tw^3&\tw^2\\
  \tw^3&\tw^2&\tw^1\\
   \tw^2&\tw^1&\tw^3\\
 \end{pmatrix},\\
&Q=\begin{pmatrix}
   0&-i & i\\
  i &0&-i\\
   -i&i&0\\
 \end{pmatrix}, 
\label{Gmatrix}
 \end{align}
where $\tw^p=w^p_3+w^p_8$, 
$A_\mu=\delta_{\mu \varphi}K_3$,  ${\hat A}_\mu=\delta_{\mu \hat n} K_{\hat n}$.
A Weyl symmetric structure of the charge matrix $G$  corresponding to interaction with vacuum gluon field
implies that matrix $G$ has three eigenvectors $\hu^{0,\pm}$ with corresponding eigenvalues
$\lambda^{0,\pm}$
\begin{align}
\lambda^0&=0,~~\lambda^{\pm}=\pm \sqrt{\tg},\nn \\
\hu^0&= 
\begin{pmatrix}
   1\\
 1\\
  1\\
 \end{pmatrix},\\
 \hu^\pm&=\begin{pmatrix}
   \tw^1\tw^3+\tw^2(\pm\tg-\tw^2)\\
 \tw^2\tw^3+\tw^1(\pm\tg-\tw^1))\\
   \tw^1\tw^2+\tw^3(\pm\tg-\tw^3)+\tg^2\\
 \end{pmatrix}, \\
\tg^2&= (\tw^1)^2+(\tw^2)^2+(\tw^3)^2-\tw^1\tw^2-\tw^2\tw^3-\tw^3 \tw^1,\nn
\end{align}
where $\tg$ is a Weyl invariant color charge.
Three vectors $\hu^{0\pm}$ form an orthogonal vector basis in the color space $R^3$. 
The vector $\hu^0$ is a common eigenvector for both charge matrices $G$ and $Q$.
Substituting quark triplet $\Psi^0=\psi(x) \hu^0$ in (\ref{GQeqns}) one obtains a free Dirac equation
for the quark mode $\psi^0(x)$
 \bea
&&   \big ( i \gamma^\mu \pro_\mu-m \big) \psi^0(x)=0.
\eea
It is surprising, the system of equations  (\ref{GQeqns}) 
admits an exact free quark solution, which is absent for the quark equation (\ref{usualeq}) obtained 
by means of the usual simple Abelian projection. This implies, that one has 
a complete basis of vacuum fields containing the Weyl symmetric Abelian gluon field $K_3$ and the quark mode
$\psi^0$ given by the vector and spinor spherical harmonics respectively.
Free gluon and quark solutions with the lowest angular momentum describe a non-degenerate color
invariant vacuum characterized by non-zero vacuum gluon and quark condensates which do not interact to each other
 in the leading order approximation.

Solutions $\Psi^\pm=\psi^\pm(x) \hu^\pm$ belong to two-dimensional Weyl representation $\Gamma_2$ 
which contains two singlet non-vector symmetry representations for two color quarks $\psi^\pm$ \cite{SM2}.
Using the Weyl symmetric Abelian projection with vanished off-diagonal fields 
one obtains the following equations for Weyl invariant quark modes $\psi^\pm$
with opposite Weyl invariant charges $\pm\tg$ 
\be
\big[\big ( i \gamma^\mu \pro_\mu-m \big)\pm g\tg \gamma^\mu A_\mu^3 \big ] \psi^\pm=0.
 \ee
 It  is clear, that solutions $\psi^\pm$ can not be transformed into each other by Weyl transformations, and they form
 two separate Weyl representations.
     We conclude, the Weyl symmetric quark solutions form an orthogonal basis 
     consisting of three singlet quarks $\Psi^{0,\pm}$ with color charges $(0, \pm \tg)$.
     Note that all three quarks $\psi^{0,\pm}$ are still color quarks since they interact at quantum level with 
     all off-shell field components of the quantum (virtual) gluon $A_\mu^a$.
 
At first glance, the existence of exact free quark solution $\Psi^0$ seems to be in contradiction with the quark confinement.
One should stress that the Weyl symmetric free gluons and quarks exist only in the classical theory. 
In quantum theory one has generation of non-trivial vacuum gluon and quark condensates. This changes drastically
the structure of the space of quantum states.
We show that interaction of the free gluon and quark with corresponding vacuum condensates leads to 
formation of localized bound states.
To prove this we calculate first a one-loop effective action in the presence of quark condensate \cite{SM2}
and demonstrate generation of a non-trivial vacuum. 
Our results imply the following expression for one-loop effective potential
\bea
&& V_{eff}^{\scriptsize\mbox{1-l}}=-m|\Phi|+\dfrac{3g^2}{8 \pi^2}\Big \{-\dfrac{b}{4}{\sqrt {-b^2+4c}}\big(\pi-  \nn\\
 &&2 {\arctan \dfrac{b}{\sqrt {-b^2+4c}}}\big)-\dfrac{c_g^2\Phi^2 }{y_0}+ \nn \\
&& \dfrac{1}{4}\big ((b^2-2 c) \log (\dfrac{c}{m^4})+2 y_0^2 \log (\dfrac{-y_0}{m^2}) \big)\Big \}, \label{Veffvac}
\eea
where
\bea
b&=&y_0+m^2, ~~c=-\dfrac{\Phi^2}{y_0},~~\Phi={\displaystyle <}0|{\bar \Psi}^0\Psi^0|0\,{\displaystyle >},\nn
\eea
and $y_0$ is a real root to cubic equation
\bea
c_g^2 \Phi^2-2c_g m \Phi y +m^4 y^2+y^3=0,\nn
\eea
 \begin{figure}[h!]
\centering
\includegraphics[width=64mm,height=42mm]{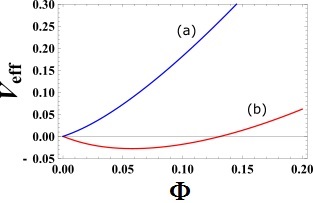}
\caption[fig2]{(a) The vacuum energy dependence on positive values of quark condensate 
$\Phi\equiv{\displaystyle <}0|{\bar \Psi}^0\Psi^0|0\,{\displaystyle >}$ (blue curve);
(b) vacuum energy dependence on negative values of the quark condensate
$\Phi\equiv-{\displaystyle <}0|{\bar \Psi}^0\Psi^0|0\,{\displaystyle >}$ (red curve).
 (in units of quark mass). 
}
\label{Fig2}
\end{figure}
 where the group number factor $c_g$ takes values $\{\dfrac{16}{9},~\dfrac{3}{2},~ \dfrac{1}{2}\}$ for gauge groups
 $SU(3)$, $SU(2)$ and $U(1)$ respectively.
The vacuum energy dependence on positive and negative values of vacuum quark condensate is depicted in Figs. 2(a), 2(b).
It is clear that a non-trivial vacuum is generated at a finite negative value of the vacuum quark condensate.
For small values of the quark condensate, $|\Phi|{\displaystyle <}1$, the effective potential has a simple form 
\bea
V_{eff}^{\scriptsize\mbox{1-l}}\simeq m \Phi+ \dfrac{3 c_g g^2}{8 \pi^2 m^2}\Phi^2 \Big (\log \big[\dfrac{c_g g^2 |\Phi|}{m^3}\big]+
\dfrac{3}{2}\Big).  \label{Veff}
\eea
In quasiclassical approximation the quark condensate is described by a spinor spherical harmonic \cite{ahiezer}
\bea
\psi^0_{jm}&=\begin{pmatrix}
\phi(x)_{jm}\\
\chi(x)_{jm}\\
\end{pmatrix}=
\begin{pmatrix}
 a(r) \Omega_{jlm\nu}\\
i b(r)\Omega_{jl'm\nu}\\
 \end{pmatrix}.
\eea
 A negative quark condensate is caused by the relativistic component $\chi$ with $l=1$, Fig. 3.
 
It is well known and commonly accepted a simple mechanism of quark confinement 
provided by a gluon string which bounds the quark anti-quark pair in a meson. It  is much less known 
on a possible mechanism of confinement of a single quark. It was proposed that origin 
of single quark confinement is related to vacuum polarization effects \cite{CKS, CKS2}. 
We consider a possible mechanism of single quark confinement based on vacuum polarization effect
leading to generation  of a non-trivial vacuum characterized by a non-vanishing negative vacuum quark condensate. 
 
 \begin{figure}[h!]
\centering
\includegraphics[width=60mm,height=48mm]{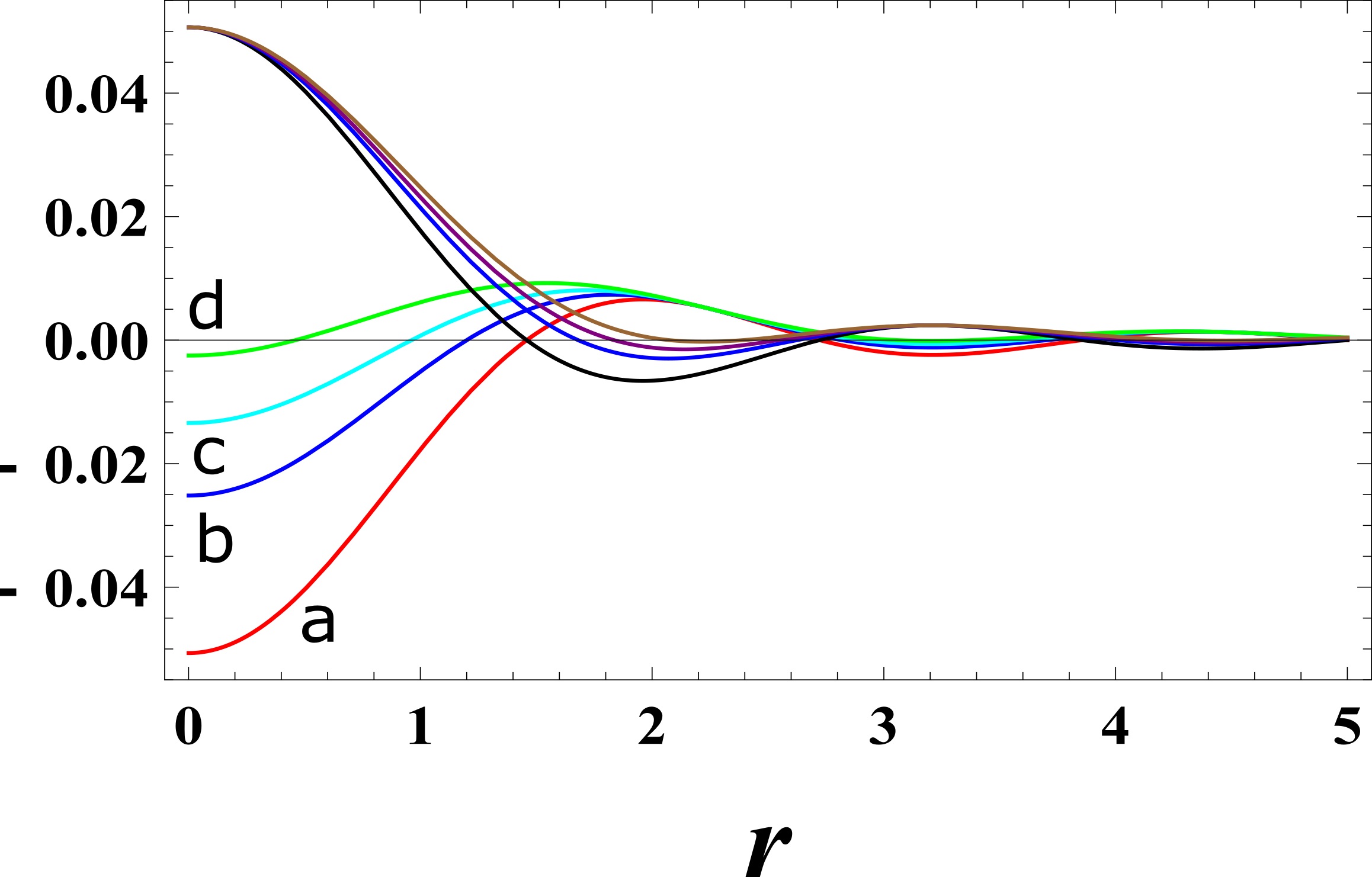}
\caption[fig3]{
Quark condensate density 
for various values of the quark mass
and orbital momentum: I. negative condensate $l=1$: (a) $m=0$, (b) $m=0.5$, (c) $m=1$, (d) $m=3$;
II. positive condensate $l=0,$ $m=(0,0.5,1,3)$ (curves from bottom to top), the dependence on mass is weak.}\label{Fig3}
\end{figure}

A single quark in such a vacuum interacts to vacuum quark condensate and forms localized bound states.
One can estimate the energy spectrum of bound states by solving
a quantum mechanical problem defined by the Dirac equation
for a quark placed in a potential well $V$ corresponding to the quantum effective potential $V_{eff}^{\scriptsize\mbox{1-l}}$
\bea
&&i \dfrac{\pro \psi}{\pro t}=(H+V) \psi, 
\eea
where  $H=\vec \alpha \vec p+\beta m$ is a Dirac Hamiltonian.
For qualitative estimate we approximate the effective potential
by a ``rectangular'' spherically symmetric well 
\bea
V(r) =\left\{{~~~~0~~, r>r_0~~;
\atop -V_0~~, r<r_0,}\right. 
\eea
where the width $r_0$ of the well is defined by hadron size, $r_0\simeq 1fm$. The depth $V_0$ is 
a free parameter since one loop approximation provides only a lower bound $ V_0\simeq 0.5 m$.
Such a relativistic quantum mechanical problem has been solved in the analytical form in quantum electrodynamics 
in a case of the Dirac  electron (\cite{ahiezer},section 1.5).
As a result, for values of the potential depth less than a critical value, $V_0 \leq V_{cr}\simeq 3.5 m$,
one has a discrete energy spectrum for bound states with an exponentially decreasing radial wave function
$\psi=\exp[-\sqrt{m^2-\epsilon^2}\,r]$ ($r\geq r_0$) and discrete energy levels $|\epsilon|\leq m$. The ground state
with orbital quantum number $l=0$ describes $S$-state with zero angular momentum which can be treated as a meson state. 
In our consideration we did not take into account back reaction of the vacuum condensate to interacting single quark.  
We assume that after creating from the vacuum a single quark with baryon, electric and fermion
charges, the condensate state has been changed taking opposite quantum numbers, so that a final bound state 
manifests quantum numbers of meson. This reminds the soft (long-time scale) mechanism of quark confinement proposed 
by Casher-Kogut-Susskind \cite{CKS, CKS2} where quarks are present in deep-inelastic processes as free point-like particles,
whereas on a longer time scale the polarization effects prevent the appearance of quarks in the final physical state. 

Note that, a single quark can not be freed from the bound state, since the quark generates a non-trivial vacuum
condensate where ever it goes , in other words, the quark stays permanently inside the potential well. 
For values of $V_0$ larger than critical value $V_{cr}$ the bound states become unstable causing creation of 
quark-antiquark pairs (\cite{ahiezer}, section 1.5). In this case  the solution turns outside the framework 
of one particle quantum mechanical problem.

\section{Localization of a singlet quantum gluon state}

Now we consider interaction of a single gluon with vacuum gluon condensate.
For qualitative estimate we apply a known expression for one-loop effective Lagrangian of $SU(3)$ QCD 
\cite{flyvb,mpla2006} 
\bea
 {\cal L}^{\scriptsize\mbox{1-l}} =-\dfrac{1}{4} \tF^2-k_0 g^2 \tF^2 \bigg(\log \Big(\dfrac{g^2 \tF^2}
{\Lambda_{\rm QCD}^4}\Big)-c_0 \bigg),
\label{Leff}
\eea
where $\tF^2 \equiv \tF_{\mu\nu}^2$ is a squared Abelian field strength of magnetic type, and we treat
 $k_0, c_0$ as free parameters.
A corresponding effective potential 
$V^{\scriptsize\mbox{1-l}}=-{\cal L}^{\scriptsize\mbox{1-l}}$ has an absolute 
minimum at positive vacuum gluon condensate value
\bea 
g^2 B_{0\,\mu\nu}^2 =\Lambda_{\rm QCD} \exp \Big(c_0-1-\dfrac{1}{2k_0g^2}\Big).\label{npcond}
\eea
We split the field strength $\tF_{\mu\nu}$ into two parts
\bea
\tF_{\mu\nu}=B_{\mu\nu} + F_{\mu\nu}, 
\eea
where $B_{\mu\nu}=\pro_\mu B_\nu-\pro_\nu B_\mu$ describes the gluon condensate,
and $F_{\mu\nu}=\pro_\mu A_\nu-\pro_\nu A_\mu$ contains the Abelian gauge potential of a single gluon. 
Decomposing the Lagrangian ${\cal L}^{\scriptsize\mbox{1-l}}$
around the vacuum condensate field $B_{\mu\nu}$ and using relation (\ref{npcond}), one obtains
an effective Lagrangian for a gluon in the field of vacuum gluon condensate
\bea
{\cal L}_{\rm eff}^{(2)}[A]=-2 k_0 g^2 \dfrac{(B^{\mu\nu} F_{\mu\nu})^2}{B^2} \equiv
-\kappa (B^{\mu\nu} F_{\mu\nu})^2 , \label{LagrBf}
\eea
where we neglect terms corresponding to the absolute
value of the vacuum energy, and $\kappa$ is a free parameter.
The effective Lagrangian ${\cal L}_{\rm eff}^{(2)}[A]$ is strikingly different from the classical Lagrangian of QCD. 
Note that this result is model independent, we could  start with Ginsburg-Landau type Lagrangian and obtain the same result.
The gauge potential of the vacuum gluon condensate is given by a magnetic vector harmonic 
$\vec A_{l=1,m=0}^{\mathfrak{m}}$, or explicitly,
\bea 
B_\varphi(r,\theta,t)=b(r)\sin^2 \theta \sin(t),
\eea
where $b(r)=r j_1(r)$.
A single gluon is described by the Abelian potential 
which assumed to be time-coherent to the gluon condensate function with a phase shift $\phi_0$,
\bea
A_\varphi(r,\theta,t)=a(r,\theta) \sin(t +\phi_0).
\eea
A time-averaged effective Lagrangian ${\cal L}_{\rm eff}^{(2)}[A]$ 
leads to Euler equation for gluon field $a(r,\theta)$ which represents 
a second order partial differential equation. It turns out that equation for a gluon interacting to vacuum condensate 
is separable for a ground state with a zero angular momentum, $j=0$. So the solution is spherically symmetric,
$a(r,\theta)=f(r)$, and satisfies a simple ordinary differential equation
\bea
r^2 b'^2 f'' -2b'(b+r b'-r^2 b'')f'& \nn \\
-\Big (r^2 b^2+\xi (-2 b^2+r^2 b'^2+r^2 b b'')\Big ) f&=0, \label{eqf}
\eea

where $\xi\equiv\cos(2\phi_0)/(2+\cos(2\phi_0))$.
The equation is solved numerically in dimensionless variables
$x=r/\nu_{11}$, $\nu_{11}=2.74\cdots$ is the first antinode of the radial function 
$b(r)= r j_1(r)$. The numeric solution implies localization of the single gluon 
inside a sphere with radius $r_0=\nu_{11}$ (or $x=1$), Fig. 4 (a). 

 \begin{figure}[h!]
\centering
{\includegraphics[width=70mm,height=52mm]{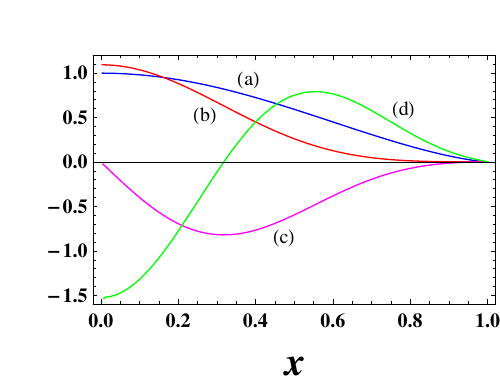}}
\caption[fig4]{(a) Solution $f(x)$; (b) a radial energy density 
$\bar{\cal E}/4 \pi \kappa$(in red); 
(c) the first derivative of the energy density; (d) the second derivative of the energy density; 
$\phi_0=\pm \pi/2$. }\label{Fig4}
\end{figure} 
The point $r_0=\nu_{11}$ represents a removable singularity, and the 
energy density has smoothly vanishing first and second radial derivatives at $r_0$, Fig. 4 (c, d).
The solution has a minimal energy at phase shift value $\phi_0=\pi/2$, and only at this value
the time averaged effective Lagrangian vanishes, as in a case of Maxwell Lagrangian for photon plane waves,
providing stability of the ground state. The ground state can be treated as a state of the lightest scalar glueball.

\section{Spectrum of lightest glueballs}

Explicit analytical expressions for quantum stable Abelian solutions are given in terms of vector spherical harmonics. 
This allows to perform standard quantization of Abelian fields defined in a finite space 
region constrained by a sphere of radius $a_0$ corresponding to an effective
glueball size.
It is suitable to introduce dimensionless units $\tilde M=M a_0, x=r/a_0, \tau=t/a_0$.
To find proper boundary conditions we require that the Pointing vector 
$\vec {\mathbf S}=\vec {\mathbf E} \times \vec {\mathbf B}$  
vanishes on the sphere. This implies two possible types of boundary
conditions \cite{MIT1}
\be
\begin{array}{rcrclcrcl}
({\rm I}): &\quad& \vec A_{lm}^{\mathfrak{m,e}} (\tilde M x)\big|_{x=1}\!\!&=&\!\!0, &\quad& \tilde M_{nl}\!\!&=&\!\!\mu_{nl},\\ [2\jot]
({\rm II}): &\quad& \pro_r(r \vec A_{lm}^{\mathfrak{m,e}} (\tilde M x))\big|_{x=1}\!\!&=&\!\!0, &\quad& \tilde M_{nl}\!\!&=&\!\!\nu_{nl}, 
\end{array}
\label{munu}
\ee
where $\tM_{nl}$ stands for nodes $\mu_{nl}$ or antinodes $\nu_{nl}$ of the Bessel function $ j_l(r)$ ($n=1,2,3,...; l=1,2,3,...$).
We choose the following normalization condition for the
vector harmonics $\vec A_{lm}^{\mathfrak{m, e}}(\tM x)$ 
\bea
\dfrac{1}{4\pi}\int_0^1\!\!{\rm d}x \!\!\int\!{\rm d}\theta\, {\rm d}\varphi\,  x^2 \sin\theta(\{\vec A_{lm}^{\mathfrak{m, e}}(\tM x))^2
=\dfrac{1}{\tM_{nl}}. \label{normcond}
\eea
The standard canonical quantization results in the following Hamiltonian
expressed in terms of the creation and annihilation operators   $c^{\pm}_{nl}$
\bea
H= \dfrac{1}{2} \sum_{n,l,m} \tilde M_{nl} ( c^+_{nlm} c^-_{nlm}+ c^-_{nlm} c^+_{nlm}). \label{Hamvac}
\eea
One particle states $\{c^+_{nlm} |0\rangle\}$ describe free gluons
which are not observable quantities since we have to take into account 
their interaction to vacuum gluon condensate. 

We apply a simple model based on one-loop effective Lagrangian of QCD which
describes appearance of localized solutions corresponding to the lightest glueballs.
Certainly, one loop effective potential is not a much appropriate tool for quantitative  description of glueballs,
nevertheless, it contains a non-perturbative part originated from summation of contributions from infinite number
of one-loop quantum corrections. This provides qualitative description of formation of glueballs as a result of interaction
of a single gluon with corresponding generated vacuum gluon condensate.

 \begin{figure}[h!]
\includegraphics[width=72mm,height=46mm]{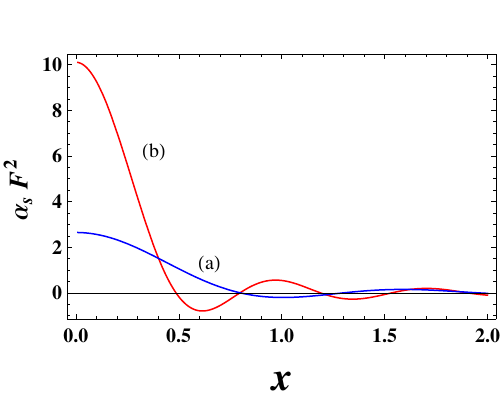}
\caption[fig5]{Radial densities of the magnetic vacuum gluon condensates $\alpha_s\langle \overline{F^2} \rangle$ 
corresponding to modes $\nu_{11}=2.74\cdots$, (a), and $\mu_{11}=4.49\cdots$, (b); $(n=l=1, m=0)$.}\label{Fig5}
\end{figure}
Qualitative estimates of the lightest scalar glueball spectrum
can be performed in a model independent way assuming that
vacuum gluon condensate is a universal order parameter for glueballs with different quantum numbers.
The knowledge of explicit solutions for the vector potential
allows to find analytical expressions for the radial density of the vacuum gluon condensate functions performing 
averaging over the time and polar angle.
Averaged over the time and polar angle vacuum gluon condensate functions
  $\alpha_s\langle (F_{\mu\nu})^2 \rangle$ corresponding to magnetic modes 
  $\vec A_{11}^{\mathfrak{m}}(\nu_{11} x)$ and  $\vec A_{ 11}^{\mathfrak{m}}(\mu_{11} x)$
    are depicted in Fig. 5 ($\alpha_s=0.5$).
The oscillating behavior of the vacuum gluon condensate density was obtained before within 
the instanton approach to QCD \cite{dorokhov1997}.
 Integrating the radial density over the interval $(0\leq x \leq 1)$ one can fit 
a value of the obtained vacuum gluon condensate parameter to the known value 
$\alpha_s\langle (F_{\mu\nu}^a)^2 \rangle=(540{\rm [MeV]})^4$,  
and obtain an explicit dependence of the glueball size on quantum number $\tM_{nl}$
\begin{align}
a_{nl}[{\rm fm}]&=\dfrac{197}{v_0} f_{c}^{1/4}(\tM) \approx \dfrac{107 \alpha_s^{1/4}}{v_0}\sqrt{\tM_{nl}},  \label{aM} \\
 f_{c}(\tM)&=\dfrac{N_{nl}^2}{12\pi \tM^2}\big((3-4 \tM^2+2\tM^4)\cos(2 \tM)-3\nn \\
&-2 \tM^2+2 \tM (3-\tM^2) \sin(2 \tM)+4 \tM^5 {\rm si} (2 \tM)\big),
\nn
\end{align}
where $v_0=540[1/{\rm fm}]$,  $N_{nl}$ is the normalization factor of the vector harmonic, 
and ${\rm si} (2\tM)$ is the sine integral function.
With this one can find the energy spectrum of light scalar glueballs $J^{PC}=0^{+ +}$
\bea
E_{nl}[{\rm MeV}]=\tk v_0 f_c^{-1/4} \tM_{nl}\approx
 \Big(\dfrac{80}{7 \alpha_s}\Big )^{1/4} \tk v_0 \sqrt {\tM_{nl}},  \label{EnM}
\eea
where $\tk$ is a free model parameter which can be fixed by fitting the energy value of the lightest glueball
\cite{kochelev2009,ochs2013}. 
The energy spectrum (\ref{EnM}) agrees with the Regge theory. \\

\section{Conclusion}

 In conclusion, we have demonstrated that a space of classical dynamical solutions for gluons and quarks contains a subspace
 of Weyl symmetric dynamical solutions which realize only singlet irreducible representations of the Weyl group. 
 Such singlet solutions 
 possess intrinsic color symmetry with respect to the Weyl color subgroup, as a consequence, they have 
 special features due to more
 higher symmetry in comparison with other non-Weyl symmetric solutions. In a particular, a Weyl singlet Abelian 
 classical vacuum solution  provides a color invariant (under Weyl color transformations) non-degenerate deepest vacuum due to 
 the highest inherent symmetry. The singlet structure of solutions with Abelian dominance phenomenon imply 
 that the full space of dynamical singlet solutions and a corresponding space of one particle quantum states can be classified
 by a finite set of discrete quantum numbers $M_{n,l}=\mu_{n,l}/\nu_{n,l}, k$ $(n,l=1,2,3,...,k=0,1,2,...)$. 
 This opens a new way towards non-perturbative formulation of the Yang-Mills theory
 based on exact gluon and quark solutions. 

\acknowledgments

Authors thank A.B. Voitkiv, A. Silenko, J. Evslin,  S.-P. Kim, A. Kotikov, A. Pimikov and Ed. Tsoi for numerous
valuable discussions. This  work is supported by Chinese Academy of Sciences  
(PIFI Grant No. 2019VMA0035), National Natural Science Foundation of China (Grant No. 11575254),and by 
Japan Society for Promotion of Science (Grant No. L19559).


\begin{thebibliography}{99}
\bibitem{Mill2000} A.M. Jaffe, ``The Millennium Grand Challenge in Mathematics", 
``Notices of the AMS", June/July 2000, Vol. 53, Nr. 6, p. 652-660. 
\bibitem{wilson1974} K.G. Wilson, Phys. Rev. {\bf D10}, 2445 (1974).
\bibitem{kogut-susskind1975} J. Kogut and L. Susskind, Phys. Rev. {\bf D16}, 395 (1975).
\bibitem{polyakov77} A. Polyakov, Nucl. Phys. {\bf B120}, 429 (1977).
\bibitem{thooft81} G. 't Hooft, Nucl. Phys. {\bf B190}, 455 (1981).
\bibitem{plb2018}  D.G. Pak, B.-H. Lee,  Y. Kim, T. Tsukioka, and P.M. Zhang, Phys. Lett. {\bf B780}, 479 (2018);
Suppl. material, https://doi.org/10.1016/j.physletb.2018.03.040.
\bibitem{savv} G.K. Savvidy, Phys. Lett. {\bf B71}, 133 (1977).
\bibitem{N-O} N.K. Nielsen and P. Olesen, Nucl. Phys. {\bf B144}, 376 (1978).
\bibitem{SM2} D.G. Pak, R.-G. Cai, T. Tsukioka, P.M. Zhang, and Y.-F. Zhou, Supplemental material.
\bibitem{neeman99} Y. Ne'eman, Symm.: Culture and Science, {\bf 10}, 143 (1999).
\bibitem{jackson} J. D. Jackson, {\it Classical Electrodynamics}, Wiley, New Jersey, 1999.
 \bibitem{abeldom1} A. Kronfeld, G. Schierholz, and U. Wiese, Nucl. Phys. {\bf B293}, 461 (1987).
\bibitem{abeldom3} T. Suzuki and I. Yotsuyanagi, Phys. Rev. {\bf D42}, 4257 (1990).
 \bibitem{flyvb} H. Flyvbjerg, Nucl. Phys. {\bf B176}, 379 (1980).
 \bibitem{mpla2006} Y.M. Cho, J.H. Kim, and D.G. Pak, Mod. Phys. Lett. {\bf A21}, 2789 (2006).
\bibitem{brodsky2012} S.J. Brodsky, C.D. Roberts, R. Shrock, and P.C. Tandy,
Phys. Rev. {\bf C85}, 065202 (2012).
 \bibitem{CKS} A. Casher, J. Kogut, L. Susskind, Phys. Rev. Lett., {\bf 31}, 792 (1973).
 \bibitem{CKS2} A. Casher, J. Kogut, L. Susskind, Phys. Rev. {\bf D10}, 732 (1974).
\bibitem{ahiezer} A.I. Akhiezer, V.B. Berestetsky, {\it Quantum Electrodynamics}, 1965,  Interscience Publishers. 
 \bibitem{MIT1} A. Chodos,  R.L. Jaffe, K. Johnson, C.B. Thorn, and V.F. Weisskopf,
Phys. Rev. {\bf D9}, 3471 (1974).
\bibitem{dorokhov1997} A.E. Dorokhov, S.V. Esaibegyan, S.V. Mikhailov, Phys. Rev. {\bf D56}, 4062 (1997).
\bibitem{kochelev2009} V. Mathieu, N. Kochelev, and V. Vento, Int. J. Mod. Phys. {\bf E18}, 1 (2009).
\bibitem{ochs2013} W. Ochs, J. Phys. {\bf G40}, 043001 (2013).
\end{thebibliography}
\end{document}